\documentclass[aps,prl,twocolumn,groupedaddress]{revtex4-1}

\usepackage{amsmath,amssymb,amsbsy,graphicx,subfigure,multirow,enumitem}
\usepackage{natbib}

\pdfoutput=1
\setlist{nolistsep}
\newcommand{\diff}[3][]{\dfrac{\mathrm{d}^{#1}#2}{\mathrm{d}{#3}^{#1}}}
\begin{document}

\title{Kinetic Effects in Dynamic Wetting} 


\author{James E. Sprittles}
\email[]{J.E.Sprittles@Warwick.ac.uk}
\affiliation{Mathematics Institute, University of Warwick, CV4 7AL}

\date{\today}

%Cover Letter Ideas:
%Gas is key effect according to Josserand and Thoroddsen
%As far as author is aware, first free surface flow of NS-Boltzmann, and certainly first for moving contact line
%Combine a model for ultrathin lubrication in Mems - disk head 
%
\begin{abstract} %600 characters including spaces
The maximum speed at which a liquid can wet a solid is limited by the need to displace gas lubrication films in front of the moving contact line. The characteristic height of these films is often comparable to the mean free path in the gas so that hydrodynamic models do not adequately describe the flow physics.  This Letter develops a model which incorporates kinetic effects in the gas, via the Boltzmann equation, and can predict experimentally-observed increases in the maximum speed of wetting when (a) the liquid's viscosity is varied, (b) the ambient gas pressure is reduced or (c) the meniscus is confined.
\end{abstract}

\pacs{dwdwdwqwd}

\maketitle
%max of 3750 words in main body

Understanding the physical mechanisms determining the maximum speed $U_{max}$ at which a liquid-gas free-surface can wet a solid substrate is a fundamental problem with applications to a range of natural and technological phenomena.  For example, if drops of rain (or pesticide) spreading across plant leaves exceed $U_{max}$ then a splash is generated which reduces the retention of liquid \citep{bergeron01} whilst coating processes must operate below $U_{max}$ in order to avoid product-destroying gas entrainment \citep{weinstein04}.

Despite the small gas-to-liquid density and viscosity ratios ($\rho_g/\rho_l \approx 10^{-3}$ and $\mu_g/\mu_l \approx 10^{-2}$ for air-water), the importance of gas dynamics has been established both in coating flows and in impact events, for the collisions of solid bodies with liquids \citep{duez07} and liquid drops with solids. In particular, recent activity in liquid drop impact has been aimed at understanding the gas' role using novel experimental techniques, see \cite{josserand16} and references therein.  Whilst a full characterisation of drop splashing remains an open problem, recent experiments in \cite{liu15} highlight the critical role of gas films durin the wetting phase.

The gas' dynamics become relevant when thin films are formed that generate lubrication effects, with experimental observations in both coating \cite{marchand12} and drop impact \cite{driscoll11} showing that the height $h$ of these films is in the range $\approx 1-10 \mu m$ as $U_{max}$ is approached \footnote{Orders of magnitude larger than the characteristic scales for roughness of the substrate.}.  At atmospheric pressure $P_{atm}$ ($atm$ will denote atmospheric values), the mean free path in the gas $\ell$ is $\approx 0.1 \mu m$, so that the Knudsen number $Kn=\ell/h\approx 0.01-0.1$.  Consequently, it has been suggested \citep{marchand12,sprittles15_jfm} that kinetic effects in the gas should be built into models for moving contact line phenomena.  

Recent models \citep{marchand12,sprittles15_jfm} account for kinetic effects by allowing for `slip', i.e. a jump in the velocity tangential to the boundary, at the gas-liquid and gas-solid interfaces, with a slip length proportional to $\ell$ \footnote{The `correction factor' $r$ in equation (10) of the Supplementary Material of \cite{marchand12}, which accounts for kinetic effects, has precisely the form of a slip model.} and the usual equations of hydrodynamics remaining in the bulk.  These models can qualitatively explain experimental observations in coating \cite{benkreira10} and drop impact \cite{xu05}, that reductions in the ambient gas pressure $P$ can suppress gas entrainment and splashing \footnote{Impact is more complex as compressibility and kinetic effects are both expected to play a role \cite{josserand16}.}:  as $\ell = \ell_{atm}P_{atm}/P$ increases with reduced $P$, slip is enhanced and gas is more easily removed from the thin film.

Research in kinetic theory has established that these `first-order' slip models are only accurate for $Kn\lesssim0.1$.  Technically, they can be derived from the Boltzmann equation for small $Kn$ \citep{cercignani00}.  Physically, they represent the case where the non-hydrodynamic effects are confined to a boundary layer of width $\approx \ell$, the so-called Knudsen layer,  which is small relative to the channel height ($\ell\ll h$) so that this additional physics can be incorporated into boundary conditions.  These models are on the edge of their applicability for dynamic wetting at atmospheric pressure, where $Kn\approx 0.01-0.1$, so that when $P$ is reduced they will be outside their limits of validity. This has been confirmed in \cite{sprittles15_jfm} where it has been shown that the situation is even more severe, as decreases in $P$ also lead to reductions in $h$, so that $Kn$ can easily exceed unity in experimentally-realisable conditions.

In this Letter, methods originally developed to predict rarefied lubrication flows in MEMS \cite{cercignani06} are used to derive a dynamic wetting model which is valid for all $Kn$.  As demanded by the physics, this model combines kinetic theory in the gas film described by the Boltzmann equation with hydrodynamics in the liquid phase governed by the Navier Stokes equations.
%
% in Marchand shows that effective viscosity uses a slip model with r= 1/(1 + 3*2.4*Kn) fitted from Andrew and Harris, where 'r' is a 'correction factor'.
%%%%%%%

\begin{figure}
     \centering
\includegraphics[width=0.95\columnwidth]{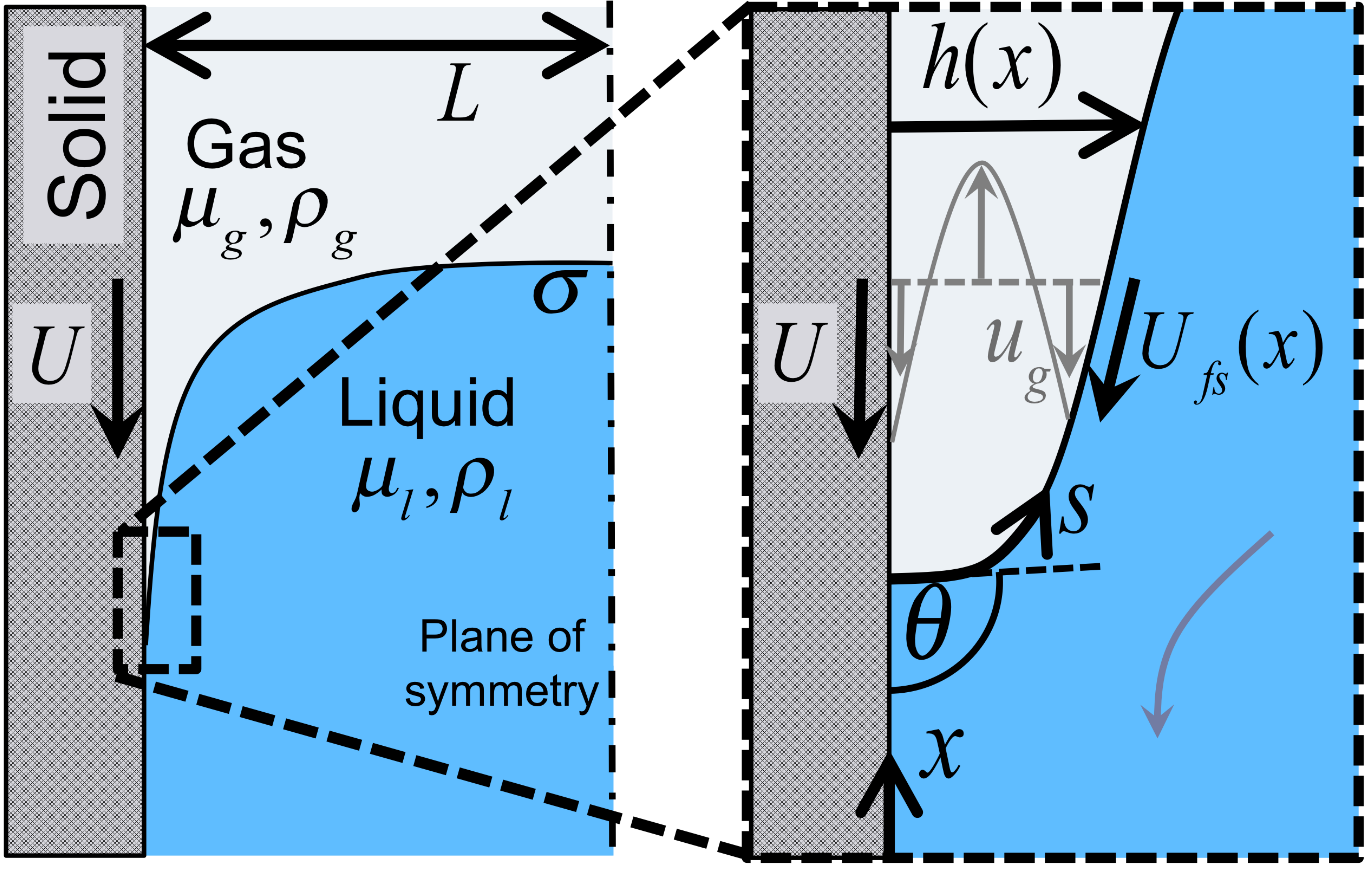}
\caption{Flow configuration (left) and close up of the thin film region (right) showing that (a) the height $h(x)$ and speed in the liquid $U_{fs}(x)$ vary with $x$, (b) there is a Couette-Poiseuille flow profile $u_g$ in the gas, and (c) that the free surface bends to attain its contact angle $\theta$. At the upper and lower boundaries of the domain, a distance $4L$ apart, is a stationary solid.}
\label{F:sketch}
\end{figure}

\emph{Flow configuration.} - The steady dynamic wetting geometry in Figure~\ref{F:sketch} allows us to consider both a coating flow, where a solid is continuously driven through a liquid bath whose free-surface is flattened by gravity, as well as the steady propagation of a meniscus confined to a microchannel of width $2L$.  These cases correspond, respectively, to $L\gg L_{\sigma}=\sqrt{\sigma/(\rho_l g)}$ and $L\ll L_{\sigma}$, where $L_{\sigma}$ is the capillary length with $\sigma$ the liquid-gas surface tension and $g$ the acceleration due to gravity.

The liquid's flow is described by the steady incompressible Navier-Stokes equations. At the liquid-solid interface conditions of impermeability and Navier-slip are used, which circumvent the moving contact line problem, choosing a fixed slip length of $l_s=10$~nm which is well within the range of experimentally observed values.  At the liquid-gas free-surface, the kinematic condition is combined with a balance of stress and capillarity.  The contact angle at which the free-surface meets the solid is assumed to be a constant $\theta_e$.  By construction, the simplest possible dynamic wetting model (formulated mathematically in the Supplementary Material) has been chosen to allow us to focus attention on the dynamics of the gas without additional parameters coming into play.  Having established the importance of the gas, more complex models for the wetting process, such as dynamic contact angles, reviewed in \cite{blake06}, can be built on top of this basic model.

As the gas flow is only strong enough to effect the liquid when it is thin, the lubrication equations can be used to describe its dynamics, see \cite{vandre_phd}.  Simulations comparing results from this formulation to full computations of the gas phase confirm the accuracy of this approach (Supplementary Material) and indicate that the gas only influences the liquid through the pressure term in the normal stress boundary condition; one may expect for $\mu_g/\mu_l \ll1$ the gas' contribution to the tangential stress condition is negligible compared to the liquid's.  Consider then, in the lubrication framework, three different models for the gas phase:
\begin{itemize}
\item \emph{No slip}: conventional model, with a non-zero slip length (fixed at $10$~nm) at the solid boundary only to circumvent the moving contact line problem.
\item \emph{Slip}:  current state of the art, with slip at the gas-solid and gas-liquid boundaries proportional to $\ell$.
\item \emph{Boltzmann}: the model developed in this Letter, with the gas phase described by kinetic theory.
\end{itemize} 

\emph{Thin film gas dynamics.} - As the process is steady, a pressure-driven Poiseuille flow forms to remove the gas dragged into the the contact line region by a boundary-driven Couette flow, caused by the motion of the solid moving at constant speed $U$ and the liquid at $U_{fs}(x)$ tangential to the free surface (Figure~\ref{F:sketch}). Such arguments are routine in hydrodynamics, formalised in the Reynolds equation, but more recently have been generalised for the Boltzmann equation \citep{fukui88,cercignani06}.  There, it is possible to identify Poiseuille and Couette flow components, but the Boltzmann equation must be solved to evaluate the respective contributions to the mass flux.  

\begin{figure}
     \centering
\includegraphics[width=1.1\columnwidth]{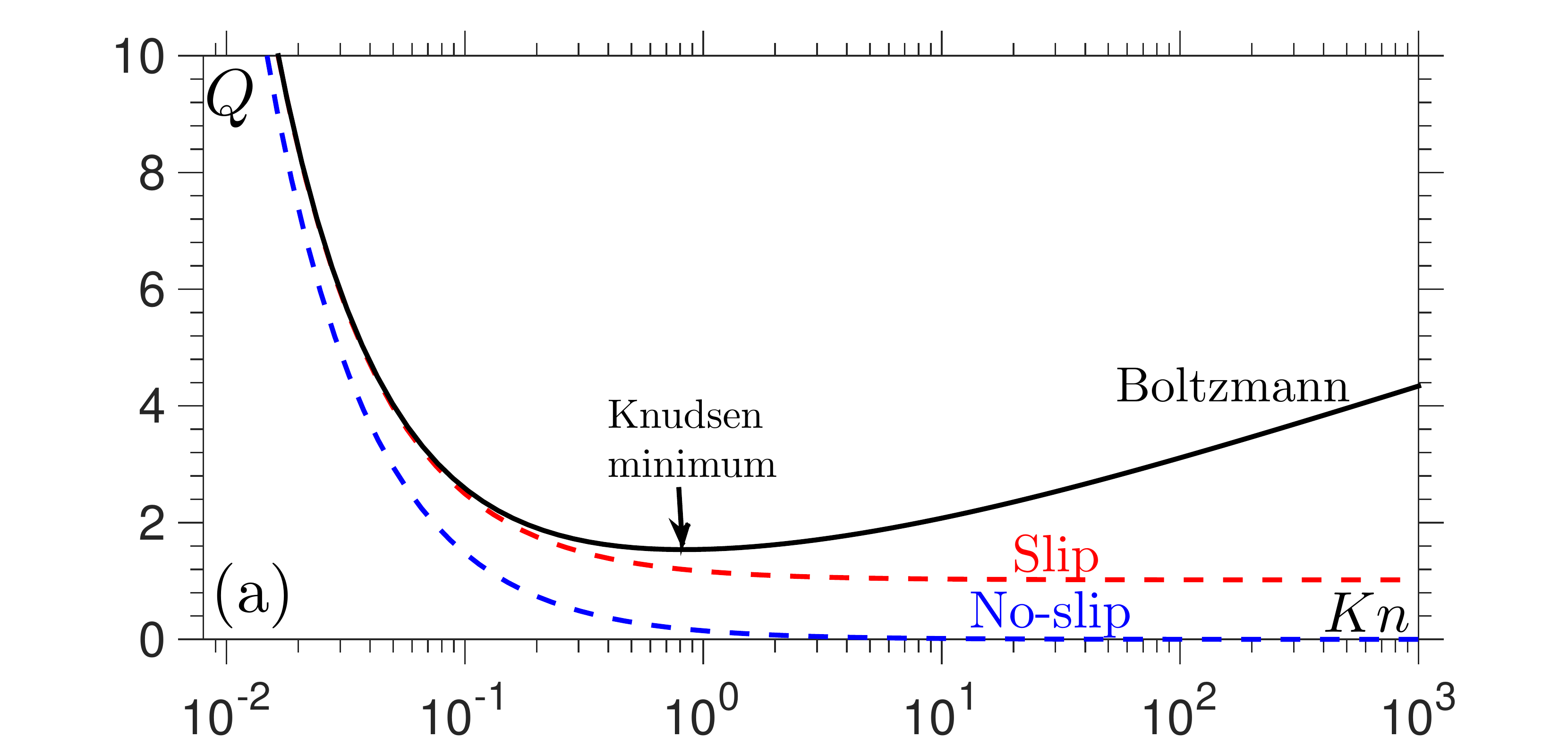}
\includegraphics[width=1.1\columnwidth]{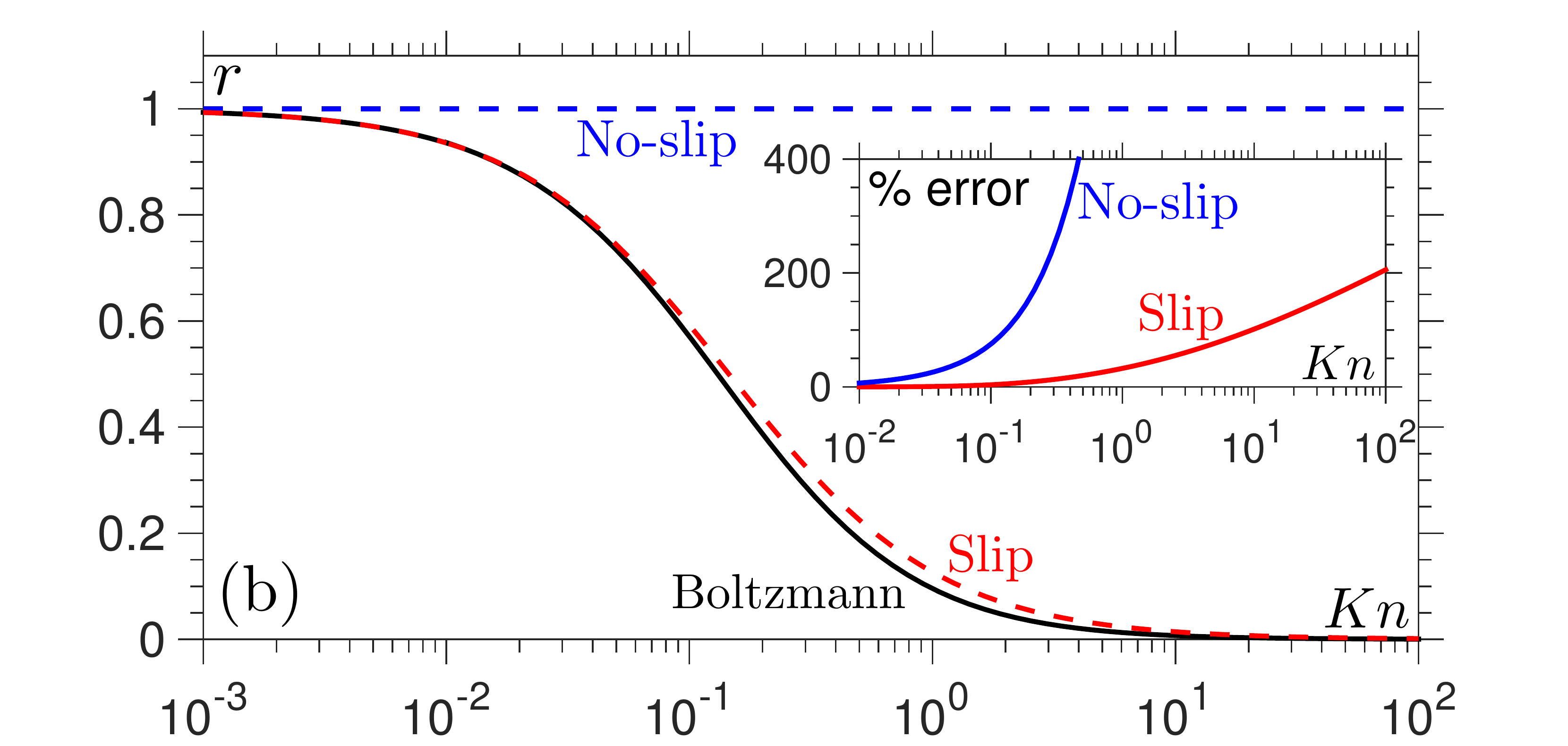}
\caption{(a) Predictions of the flow coefficients $Q(Kn)$ from (\ref{m_C}). (b) Curves of $r$, defined in (\ref{thinfilm}), and demonstration (inset) that the slip model diverges from the Boltzmann solution for larger $Kn$.}
\label{F:Qandr}
\end{figure}

Assuming diffuse reflection of molecules from each boundary, which is a sensible starting point, due to symmetry the mass flux from the Couette flow $m_C$ remains the same for all models whilst the plane Poiseuille flow contribution $m_P$ is model-dependent 
\begin{equation}\label{m_C}
m_C = \frac1 2 \rho_g h (U+U_{fs}), \quad m_P = \frac{\rho_g h^2 \ell}{\sqrt{\pi}\mu_g} \diff{p}{x} Q(Kn),  
% = - \frac{h^2}{\sqrt{2\hat{R}\hat{T}}} \diff{\hat{p}}{\hat{x}} Q(Kn) using \sqrt{\pi} \hat{\mu}_g = \hat{\rho} \hat{\ell} \sqrt{2\hat{R}\hat{T}} 
\end{equation}
where $Q(Kn)$ are the so-called `flow coefficients' \citep{sharipov98} that depend on the model used (Figure~\ref{F:Qandr}a) and $p$ is the local pressure.   For this problem it is reasonable to assume incompressible flow (see Figure~\ref{F:benkreira}b for confirmation), although the extension to compressible flow is not difficult, see \cite{gopinath97}. 

Notably, only the Boltzmann equation captures the famous `Knudsen minimum' (Figure~\ref{F:Qandr}a) in the mass flux of gas through a channel of fixed $h$ (so that $\hbox{$\frac{\rho_g h^2 \ell}{\sqrt{\pi}\mu_g}$}$ is a constant) which is driven by a constant pressure gradient. 

Conservation of mass ($m_C = -m_P$) then gives
\begin{equation}\label{thinfilm}
-\frac{h^2}{12\, r \mu_g} \diff{p}{x} = \frac{U+U_{fs}}{2}, \quad r(Kn) = \frac{\sqrt{\pi}}{12 Kn\, Q(Kn)},
\end{equation}
where it is noted that $h$, $U_{fs}$ and $Kn$ all vary along the film.  The form of (\ref{thinfilm}) suggests that an effective viscosity could be defined as $\mu_g^{\mathrm{eff}} = r \mu_g$, as considered in \citep{benkreira10,marchand12}, in order to absorb kinetic corrections into a hydrodynamic framework, and this idea has recently been pursued in pioneering drop collision simulations \cite{li16a}. Notably, at $Kn=0.1, 1, 10$ it is found that $\mu_g^{\mathrm{eff}}/ \mu_g = 0.57, 0.096, 0.0071$, showing the rapid drop in resistance as the film height is reduced.  Although this gives us a picture of the role that $r$ plays, it can confuse matters, as pointed out in \cite{hadji06}, as $\mu_g^{\mathrm{eff}}$ is problem specific and dependent on the ambient pressure, whilst $\mu_g$ is not.  Therefore, $r(Kn)$ is retained in the formulation.  

For $Kn\ll1$, the Knudsen layer is small relative to the channel height and the results of the Boltzmann equation are equivalent to using the Navier Stokes equations with Navier slip at the boundary accounting for the non-hydrodynamic effects. For the \emph{slip model} $r = (1+6 \alpha Kn)^{-1}$, where $\alpha \ell$ is the slip length and $\alpha\approx 1$ is a parameter which depends on both the collision model in the Boltzmann equation and the accommodation coefficient of the surface \cite{hadji06}.  For $Kn=0$, $r=1$ \footnote{The \emph{no-slip model} has $r=(1+4l_s/h)^{-1}$ and an extra contribution $rU_{fs}l_s/h$ on the right hand side of (\ref{thinfilm}).}.

Methods for obtaining $r(Kn)$, or more typically $Q(Kn)$, from the Boltzmann equation are reviewed in \cite{sharipov98} and often involve simplifications from a BGK approximation and/or linearisation.  Experimental data is well captured by most variants and so here the simplest possible approach of linearised BGK (where $\alpha=1.15$ \footnote{This parameter is relatively insensitive to the collision model used, for example hard sphere collisions give $\alpha=1.11$}) is used. To solve this model a variational method proposed in \cite{cercignani04} is implemented, which is shown in \cite{sharipov98} to be the simplest method for accurately approximating $Q$ (giving results within 2\% of the exact Boltzmann solution).  This leads to the curves for $Q,~r$ labelled Boltzmann in Figure~\ref{F:Qandr}.  Whilst the slip model appears satisfactory at first glance of Figure~\ref{F:Qandr}b, the inset shows that the relative error of $r$ from the Boltzmann solution becomes unacceptable for $Kn\gtrsim0.1$. 

\emph{Simulations.} - The problem is solved using a multiscale finite element framework developed in \cite{sprittles12_ijnmf}, and first applied to gas entrainment phenomena in \cite{sprittles15_jfm}, where it was benchmarked with a similar code \citep{vandre12}.  As there are length scales of nanometres ($l_s$), micrometres ($\ell$) and millimetres ($L_{\sigma}$) in the problem, the computational mesh, based on an arbitrary Lagrangian Eulerian description, has to be specially designed to capture all of the physical effects.  The main output from this code is the maximum speed of wetting $U_{max}$, past which no steady two-dimensional solutions exist \citep{sprittles15_jfm} and gas entrainment is expected to occur.  As noted in \cite{snoeijer13}, predictions of flow transitions are ideal candidates for comparing models for moving contact line phenomena, as they are easily observed experimentally, in contrast to measurements of the contact angle. 

The extension of this code to allow for a thin film description of the gas flow is relatively straightforward, and as suggested in \citep{jacqmin04}, and developed in \cite{vandre_phd}, it is assumed that $dx\approx ds$ (Figure~\ref{F:sketch}) in order to circumvent regions where the thin film approximation is not strictly valid.  Benchmark simulations in the Supplementary Material show the scheme is highly accurate, giving values for $U_{max}$ that are indistinguishable from those obtained when solving the full problem in the gas.  

Values for $r(Kn)$ obtained from the Boltzmann equation could either be calculated `on the fly', i.e. when required by the code (a `concurrent' approach), or before the code is run (a `sequential' method). For simplicity, the sequential method is chosen and the Supplementary Material provides the code used to generate $r(Kn)$.

\begin{figure}
     \centering
\includegraphics[width=1.1\columnwidth]{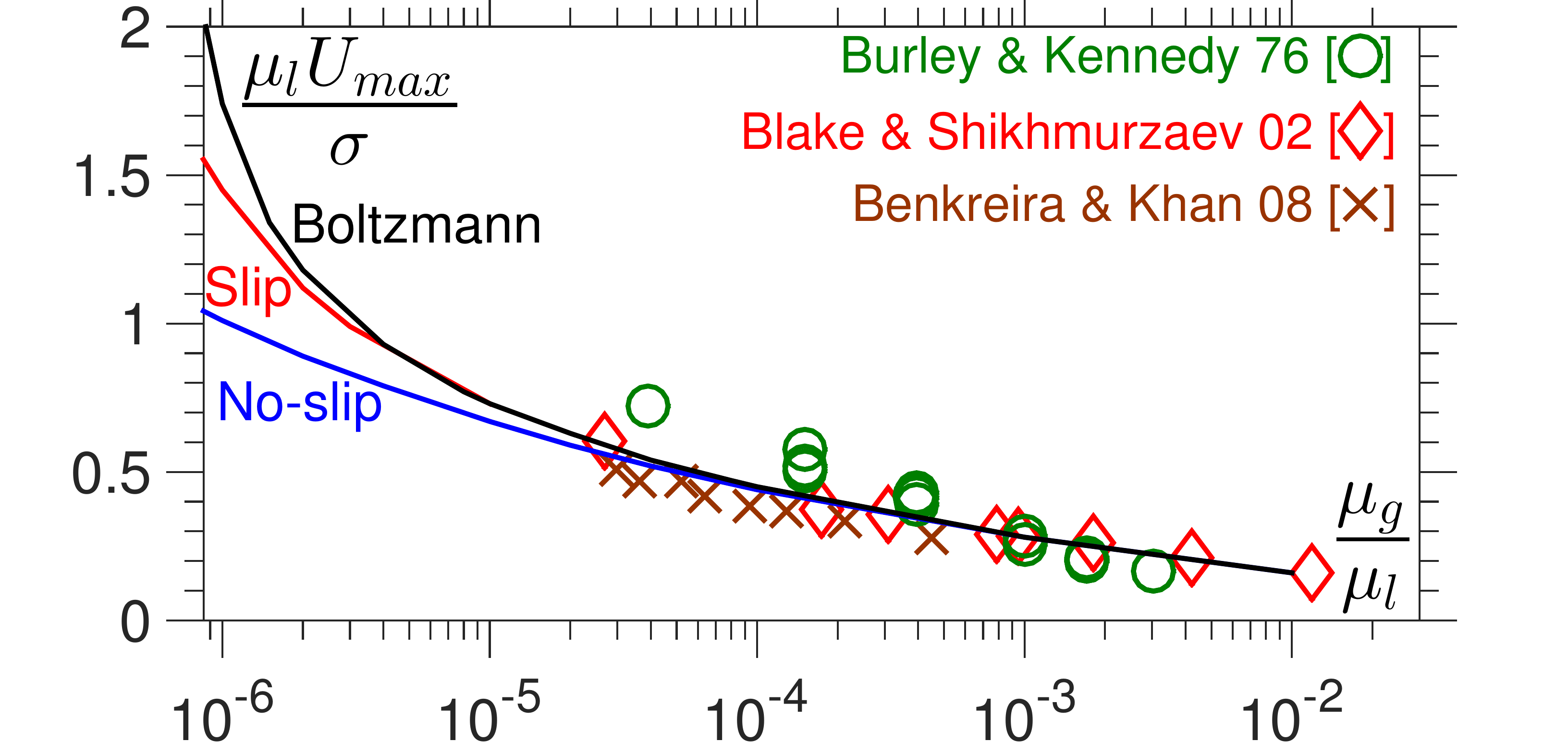}
\caption{Maximum speed of wetting (dimensionlessly a capillary number $Ca_{max}=\mu_l U_{max}/\sigma$) of water-glycerol solutions in air as a function of viscosity ratio $\mu_g/\mu_l$.} 
\label{F:viscosity}
\end{figure}

\begin{figure}
     \centering
\includegraphics[width=1.1\columnwidth]{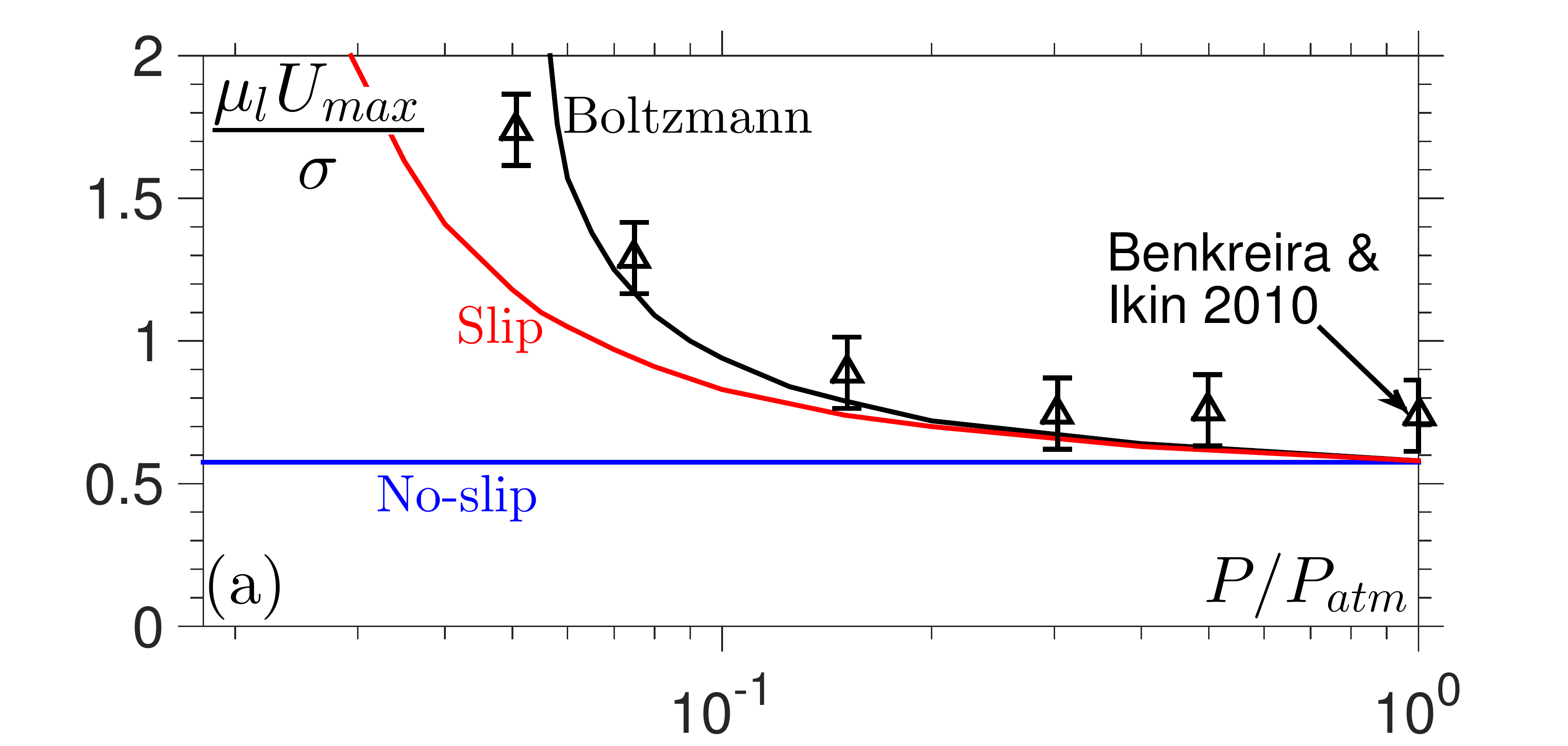}
\includegraphics[width=1.1\columnwidth]{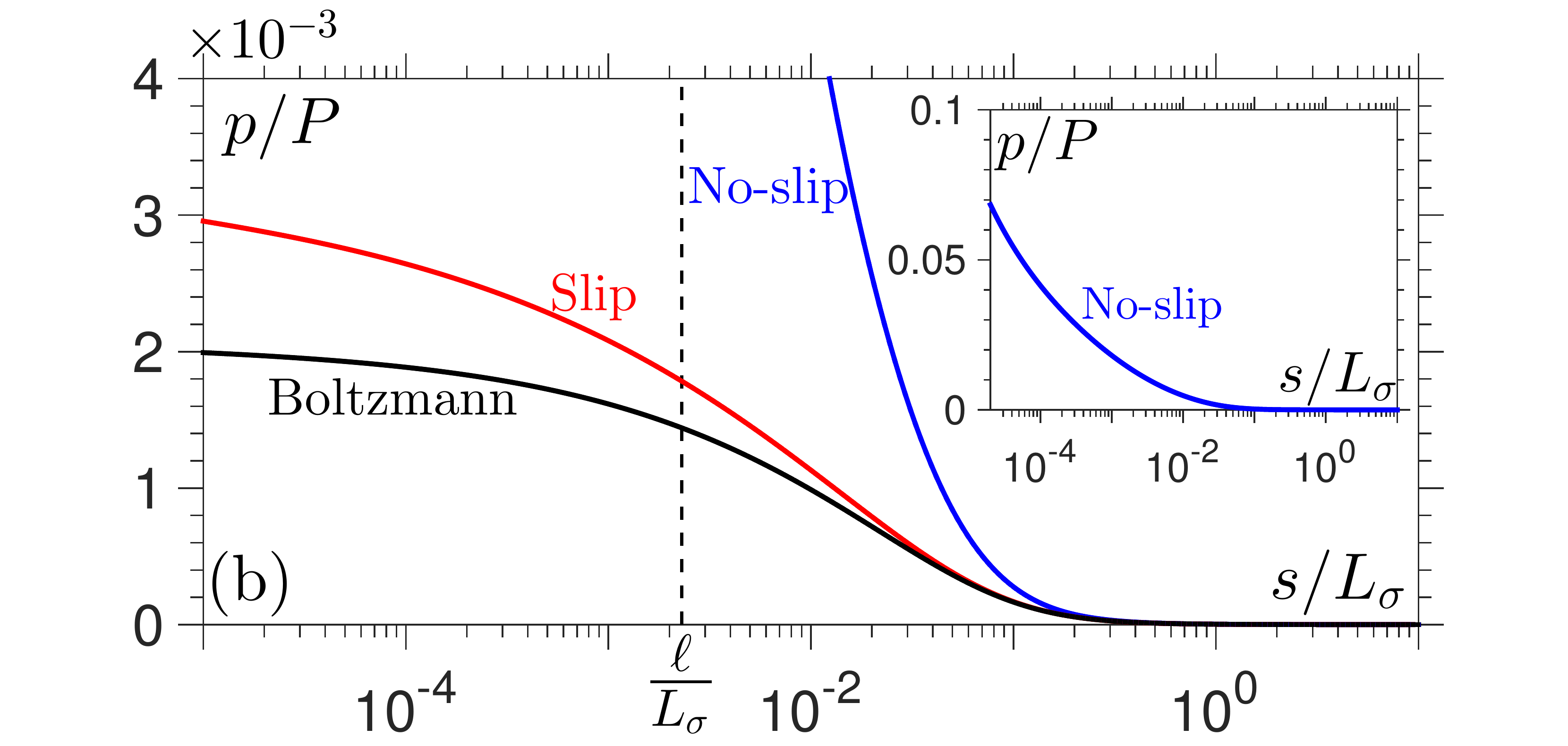}
\caption{(a) Maximum speed of wetting, for silicone oil in helium, as ambient pressure $P$ is reduced. (b) Local pressure $p$ relative to its far field value and normalised by the ambient value $P=0.06 P_{atm}$, at a contact line speed $\mu_l U/\sigma=0.5$, just before the no-slip model predicts entrainment.  The distance along the free surface $s$ starts in the thin film region.} %from values just outside the region of high curvature near the contact line %%which is still well below the mean free path $\ell/L_{\sigma}=2.3\times 10^{-3}$ at this $P$
\label{F:benkreira}
\end{figure}

\emph{Atmospheric pressure.} - Standard dip coating experiments measure the air ($\ell_{atm}=70$~nm, $\mu_g=18\mu$Pa~s) entrainment speed $U_{max}$ for different liquids on a range of solid substrates.  In Figure~\ref{F:viscosity}, this data is shown for water-glycerol solutions where $\sigma=65$~mN~m$^{-1}$ is approximately constant, so that the effect of varying the liquid's viscosity (dimensionlessly $\mu_g/\mu_l$) can be isolated.   Despite no attempt to fit the data ($\theta_e=90^\circ$ is fixed), the theoretical predictions are in good agreement with the experiments and support the validity of the approach.  

Remarkably, for $\mu_g/\mu_l < 10^{-5}$ kinetic effects become prominent at \emph{atmospheric pressure}, as the gas film's height shrinks becomes comparable to $\ell_{atm}$.  This creates a dependence on $\mu_g/\mu_l$ which diverges from the no-slip model, with the slow logarithmic increase of $\mu_l U_{max}/\sigma$ blown away by a rapid power-law-type increase.  Interestingly, this is supported by experiments in \cite{burley76} that for high viscosity liquids $U_{max}\to 0.1$~ms$^{-1}$, corresponding to $\mu_l U_{max}/\sigma \to (\mu_g/\mu_l)^{-1}$, and similar scalings are in \cite{marchand12}.  Clearly, further experimental analysis at $\mu_g/\mu_l < 10^{-5}$ is required to properly elucidate the new trends.

\emph{Reduced ambient pressure.} - Consider dip coating experiments performed in \cite{benkreira10} (see their Figure 9) with silicone oil (viscosity $\mu_l=112$~mPa~s, density $\rho_l=985$~kg~m$^{-3}$, surface tension $\sigma=17.9$~mN~m$^{-1}$, equilibrium contact angle $\theta_e=19.5^{\circ}$) as the coating liquid and helium as the ambient gas, inside a pressure-controlled chamber.  Helium's mean free path $\ell_{atm}=190$~nm at atmospheric pressure $P_{atm}=10^5$~Pa is three times larger than that of air, so kinetic effects will be enhanced, whilst its viscosity is similar $\mu_g=19\mu$Pa~s.

In Figure~\ref{F:benkreira}, the computational results are compared to experimental data.  The main results are that (a) the no-slip model completely misses the qualitative trend of $U_{max}$ as the pressure is reduced, (b) the Boltzmann model diverges from the slip model once the ambient pressure has been reduced by a factor of ten, and (c) the Boltzmann model appears to slightly better describe the experimental data. Such agreement between theory and experiment is remarkably good, when allowing for the simple dynamic wetting model implemented and the fact there are no parameters to fit.  However, the key message is that kinetic effects play a role in moving contact line phenomena in experimentally-accessible regimes.

Typical profiles in Figure~\ref{F:benkreira}b show how the model chosen changes the pressure distribution in the film.  It is clear that no-slip drastically over-predicts the pressure, peaking at $p=0.07P$ whilst the slip model's prediction is just 4\% of this value. The Boltzmann model gives further substantial reductions.  Variations in pressure along the film are consistent with gas incompressibility.  

\emph{Confined menisci.} - By considering the effect of $U_{max}$ on flow dimension $L$ (Supplementary Material), full kinetic effects are also shown to be critical for `microfluidic flow', such as for a meniscus confined to a microchannel.

\emph{Physical mechanisms.} - In \cite{vandre13}, careful simulations identified that entrainment occurs when the capillary forces at the free surface can no longer sustain the pressure gradients required to pump gas away (via a Poiseuille flow) from the contact line region.  Global balances of capillary and viscous forces have also been used in unsteady processes to predict splashing \citep{duez07,riboux14} and microdrop emission \citep{aguilar11}, where exceeding $U_{max}$ results in the contact line being left behind the advancing liquid front \footnote{Shown in \cite{riboux14} to be a sufficient, but not necessary, condition for splashing}.  

Interestingly, computations here, and in \cite{vandre13}, show that the free surface shape is relatively insensitive to the gas dynamics. Its shape is determined by the balance of viscous forces in the liquid with capillary forces at the interface. Wettability then enters the model as a boundary condition for the free surface shape, as does confinement when the channel is sufficiently narrow.  Therefore, given this profile we can compare the pressure build up in the film for the different models of the gas.

Isolating the Poiseuille flow component, equation (\ref{m_C}) shows that the pressure gradient required to drive a given mass flux $m_P$ is inversely proportional to $Q(Kn)$, so that pressure increases will be least for the Boltzmann model, where $Q(Kn)$ is largest (Figure~\ref{F:Qandr}a), as confirmed by Figure~\ref{F:benkreira}b. Physically, the increased $Q$ predicted by the Boltzmann model occurs for $Kn>0.1$ as non-equilibrium effects not only cause slip at the wall, but also drive a non-Newtonian bulk flow.  As expected, from Figures~\ref{F:viscosity} and \ref{F:benkreira}a, it is the models with the smallest increases in pressure, i.e. the largest $Q(Kn)$, which predict the largest $U_{max}$.  

Encouragingly, it appears that the free surface shape could be calculated independently of the gas phase and used to infer the maximum pressure gradients which the free surface can sustain. This is where the dependencies on the capillary number, wettability and confinement would enter the model. With this information $U_{max}$ could be calculated from lubrication theory for the gas phase, where kinetic effects would enter. However, as no simple method currently exists for characterising the required free surface profiles, a less computationally intensive model based on these ideas remains an open problem.

\emph{Discussion.} - Simulations have identified situations where conventional approaches fail to predict  $U_{max}$ due to an inadequate description of the flow physics in the gas film. Whilst the slip model captures the qualitative behaviour, the Boltzmann equation is required for quantitative predictions and thus deserves further attention. Incorporating this new physics into existing codes is relatively simple and could play a role in a wide range of free surface flows where gas microfilms appear such as in the collisions of liquid drops \citep{qian97}; the formation of tip-singularities in free-surfaces \citep{dupont06}; the stability of nanobubbles on solids \citep{seddon11}; the impact of projectiles on liquid surfaces \citep{truscott14}; and the creation of anti-bubbles from air films \citep{beilharz15}.  Furthermore, these findings motivative new directions of research, such as (a) understanding how gas molecules interact with moving liquid-gas free surfaces, from experiments or molecular dynamics simulations, and (b) developing non-lubrication formulations of the gas flow, such as moment methods approaches \cite{struchtrup05}.

The author thanks Duncan Lockerby and Alex Patronis for useful discussions and the reviewers for their constructive criticism. This work was supported by the Leverhulme Trust (Research Project Grant) and EPSRC (grant EP/N016602/1). 

% Add additional references from the Supplementary Material
\footnotetext{See Supplemental Material [url], which includes Refs.\cite{shik07,blake69,sprittles12_jcp,oron97,vandre14,cercignani66,hickey90,cercignani63}}

\bibstyle{apsrev4-1}

\bibliography{Bibliography}

\end{document}